\documentclass[twocolumn,preprintnumbers,superscriptaddress,landscape,nofootinbib]{revtex4}
\usepackage[utf8]{inputenc}
\usepackage{amsmath}
\usepackage{amsfonts}
\usepackage{amsthm}
\usepackage{float}
\usepackage{braket}
\usepackage{graphicx}
\usepackage{url}
\usepackage{tikz}
\usepackage{quantikz}
\usepackage[colorlinks=true, pdfstartview=FitV, linkcolor=red, citecolor=blue, urlcolor=blue]{hyperref}
\usepackage{slashed}
\usepackage{xcolor}

\newcommand{\beq}{\begin{equation}}
\newcommand{\eeq}{\end{equation}}

\begin{document}

\title{Signatures of baryon junctions in semi-inclusive deep inelastic scattering}

\begin{abstract}
 Local gauge invariance of the baryon wave function leads to the emergence of a baryon junction, where three (or $N$, in $SU(N)$ gauge theory) string operators merge. The existence of baryon junction dramatically affects the dynamics of baryon stopping at high energies, and the corresponding predictions are supported by the recent data from STAR Collaboration at the Relativistic Heavy Ion Collider. Here we outline the ways in which the baryon junctions can be tested in semi-inclusive deep inelasttic scattering at Jefferson Laboratory and the future Electron Ion Collider.
\end{abstract}

\author{David Frenklakh}
\email{david.frenklakh@stonybrook.edu}
\affiliation{Center for Nuclear Theory, Department of Physics and Astronomy,
Stony Brook University, Stony Brook, New York 11794–3800, USA}
\affiliation{Center for Frontiers of Nuclear Science, Department of Physics and Astronomy,
Stony Brook University, Stony Brook, New York 11794–3800, USA}

\author{Dmitri E. Kharzeev}
\email{dmitri.kharzeev@stonybrook.edu}
\affiliation{Center for Nuclear Theory, Department of Physics and Astronomy,
Stony Brook University, Stony Brook, New York 11794–3800, USA}
\affiliation{Center for Frontiers of Nuclear Science, Department of Physics and Astronomy,
Stony Brook University, Stony Brook, New York 11794–3800, USA}
\affiliation{Department of Physics, Brookhaven National Laboratory
Upton, New York 11973-5000, USA}

\author{Wenliang Li}
\email{wenliang.li@stonybrook.edu}
\affiliation{Center for Frontiers of Nuclear Science, Department of Physics and Astronomy,
Stony Brook University, Stony Brook, New York 11794–3800, USA}
\maketitle

The quark model assigns the baryon number to quarks. Quantum chromodynamics is a gauge theory, and the physical states of QCD have to be gauge invariant. A non-local quark model baryon state 
$q(x_1) q(x_2) q(x_3)$ is however not gauge-invariant, as the quark fields acquire different phases under a gauge transformation. The way to construct a gauge-invariant baryon wave function was proposed in \cite{Rossi:1977cy}, and amounts to introducing path-ordered exponential operators  (string operators) of the form 
$$P(x_n,x) \equiv {\cal P} \exp\left({i g \int_{x_n}^x A_\mu dx^\mu}\right) $$
of the gluon field $A_\mu \equiv t^a A_\mu^a$, so that the baryon state is defined as 
\begin{eqnarray}\label{baryon}
    B(x) = \epsilon^{ijk}\ \left[P(x_1,x)\ q(x_1)\right]_i\ \left[P(x_2,x)\ q(x_2)\right]_j\\ \nonumber
    \left[P(x_3,x)\ q(x_3)\right]_k .
    \end{eqnarray}
The string operator $P(x_n,x)$ acting on a quark field located at space-time point $x_n$ makes it transform under gauge transformations as a field at point $x$. The antisymmetric tensor then makes a color-singlet and gauge-invariant state out of the three quarks. 

As usual in gauge theories, imposing gauge invariance introduces dynamics. Indeed, at strong coupling the appropriately named string operators indeed describe strings connecting the quarks to the common ``baryon junction", which thus becomes a constituent of the baryon, along with the three quarks \cite{Rossi:1977cy}. 
\vskip0.3cm

As proposed in \cite{Kharzeev:1996sq}, the baryon junction in high energy collisions can be separated from the valence quarks, and the final-state baryon always emerges around the junction. In other words, it appears that the baryon number has to be assigned to the junction, and not to the quarks! Indeed, consider a high-energy inelastic proton-proton collision. The valence quarks carry a high fraction of the proton's momentum and are hard to stop -- they therefore populate the fragmentation regions of the collision. However the baryon junctions are made of gluons, carry much smaller momentum and can thus be stopped in the mid-rapidity region of the collision. When this happens, the strings connecting the junction to the valence quarks break producing a large number of quark-antiquark pairs, but the final-state baryon is always produced around the junction. It is important to note that it is uncorrelated in flavor with the original baryon, even though it carries the original junction.  
\vskip0.3cm

The energy dependence of baryon stopping observed at RHIC and LHC \cite{Brandenburg:2022hrp, Lv:2023fxk, ALICE:2013yba} confirm the predictions \cite{Kharzeev:1996sq} based on the baryon junction picture. Moreover, very recently STAR Collaboration at RHIC observed a clear difference  between the stopping of electric charge and baryon number in collisions of RuRu and ZrZr isobars \cite{Brandenburg:2022hrp}. 
\vskip0.3cm

In this letter, we argue that the baryon junction should also manifest itself in the semi-inclusive forward baryon production in electron-proton scattering. Our predictions can be tested at Jefferson Lab and, at much higher energies and a wider range in rapidity, at the future Electron Ion Collider. 
\vskip0.3cm

Note that a substantial baryon-antibaryon asymmetry at small $x$ was reported by the H1 Collaboration at HERA \cite{H1:1998ap}, and was interpreted \cite{Kopeliovich:1998ps} in terms of perturbative baryon number transfer mechanism with intercept of 1 \cite{Kopeliovich:1988qm}. This mechanism is however inconsistent with RHIC and LHC data on baryon stopping \cite{Brandenburg:2022hrp, Lv:2023fxk, ALICE:2013yba} that clearly indicate the decrease of stopping with energy and dependence of baryon number distribution on rapidity.
\vskip0.3cm

\begin{figure*}
\includegraphics[scale=0.12]{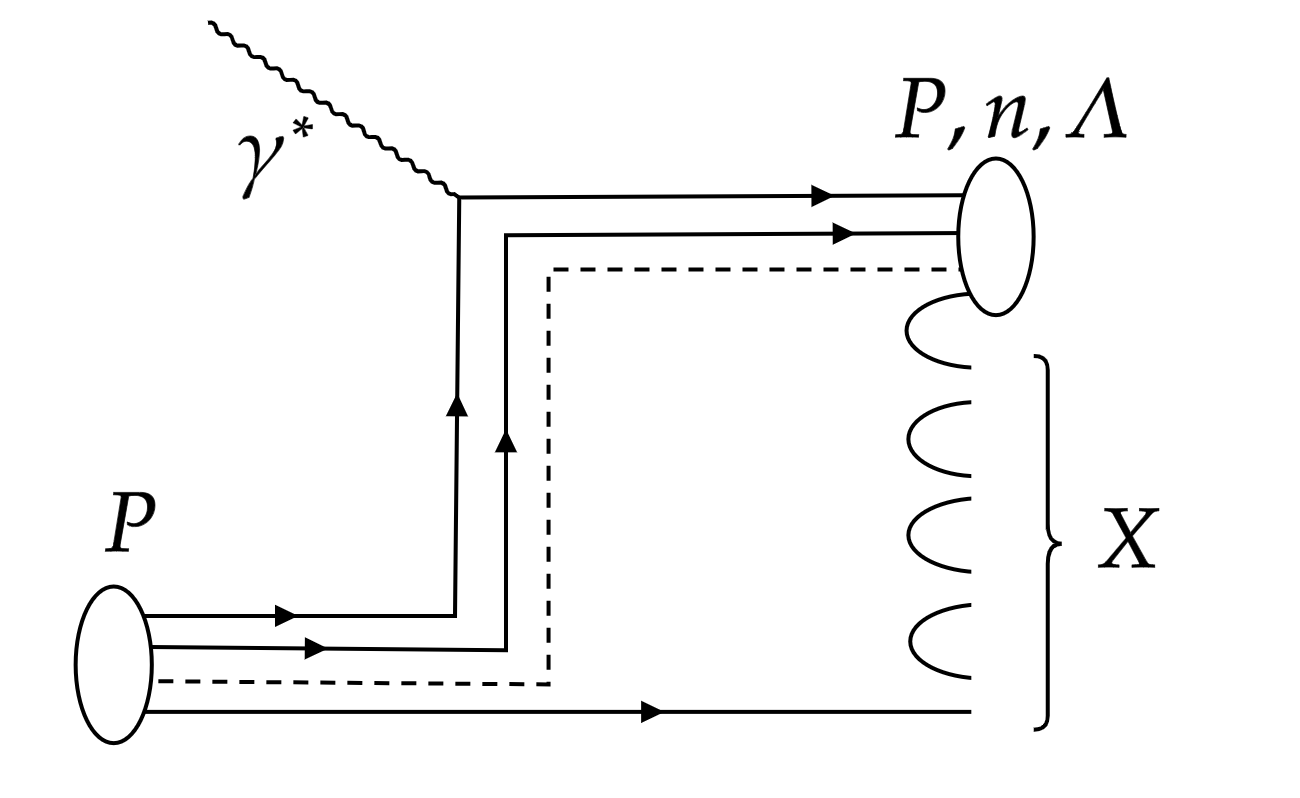} 
\includegraphics[scale=0.12]{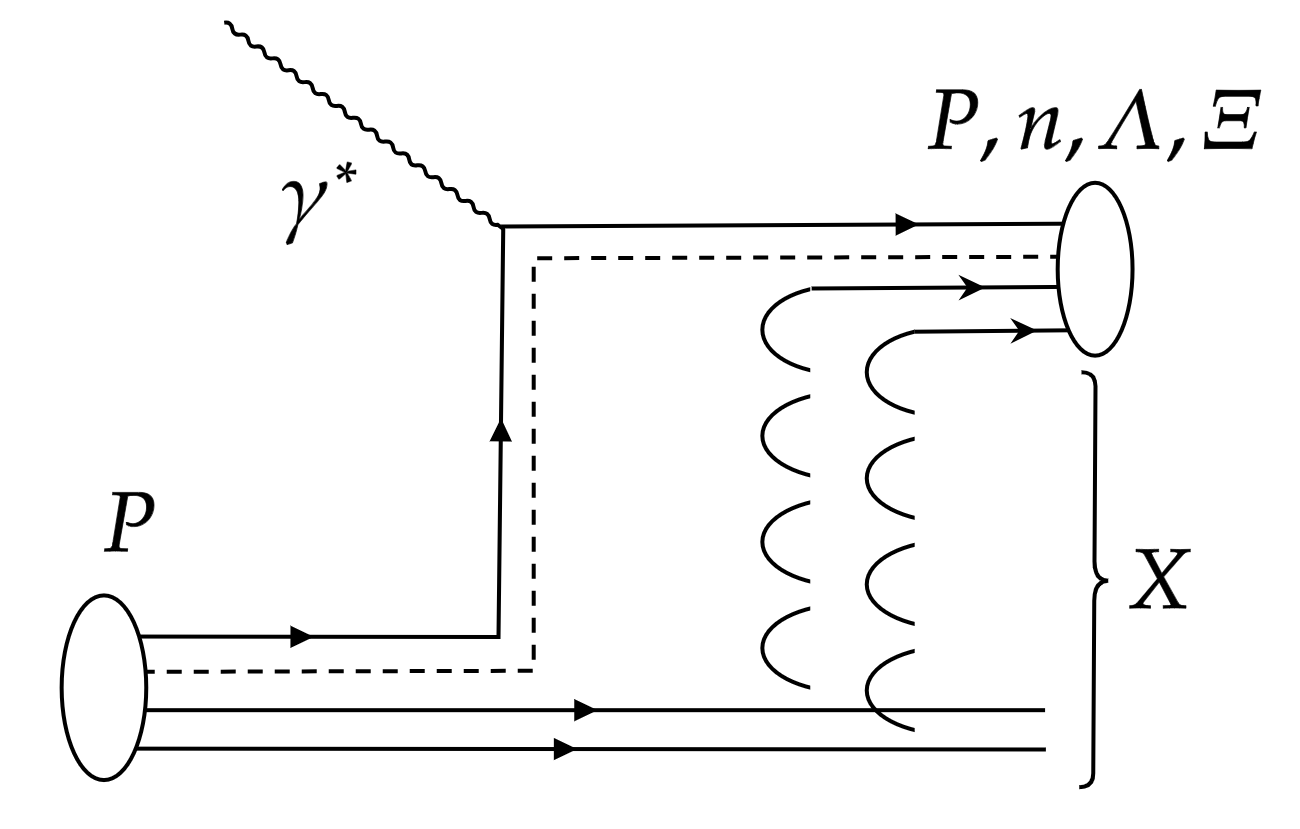} 
\includegraphics[scale=0.12]{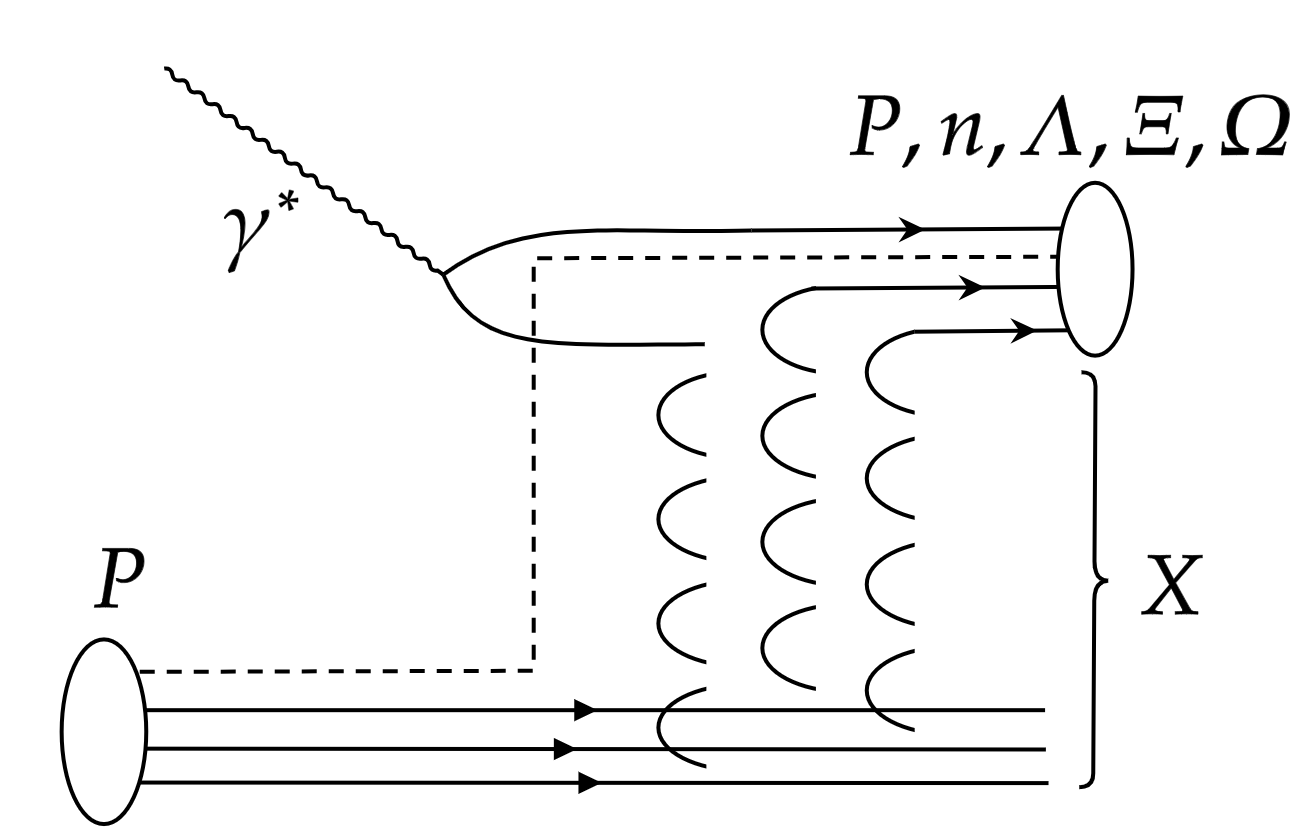} 
\includegraphics[scale=0.11]{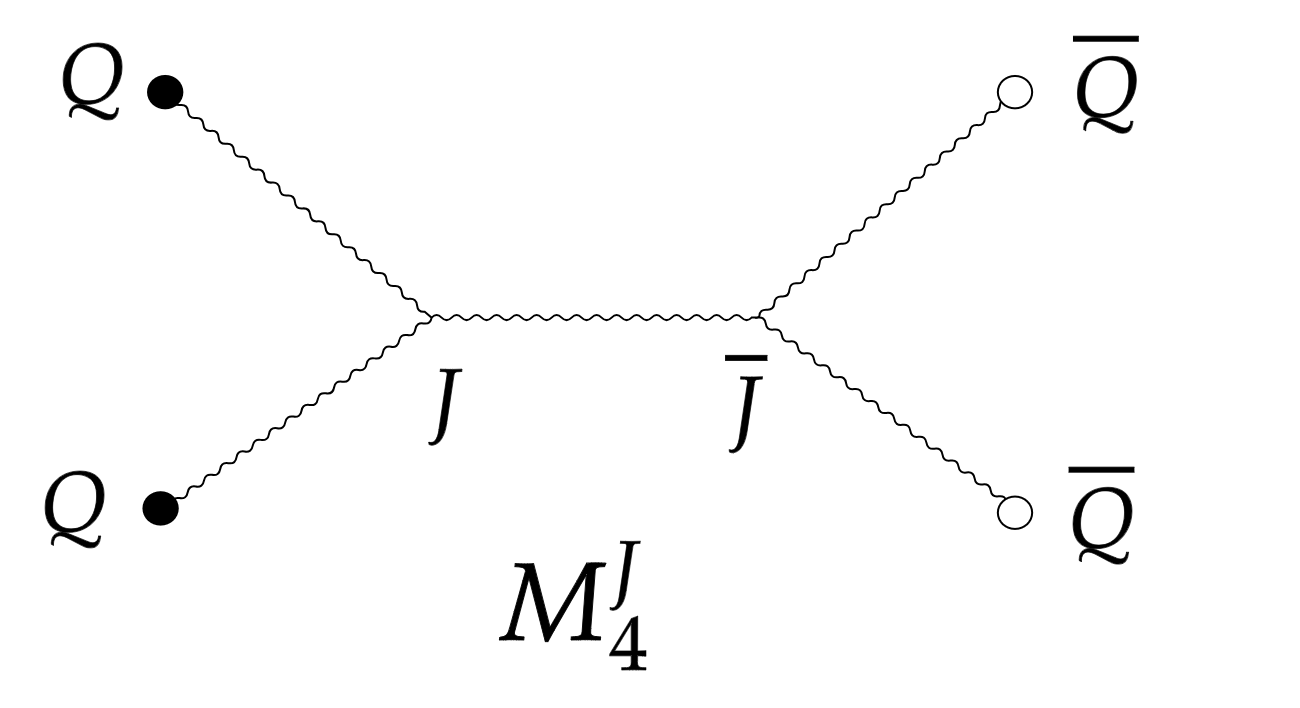}
\includegraphics[scale=0.11]{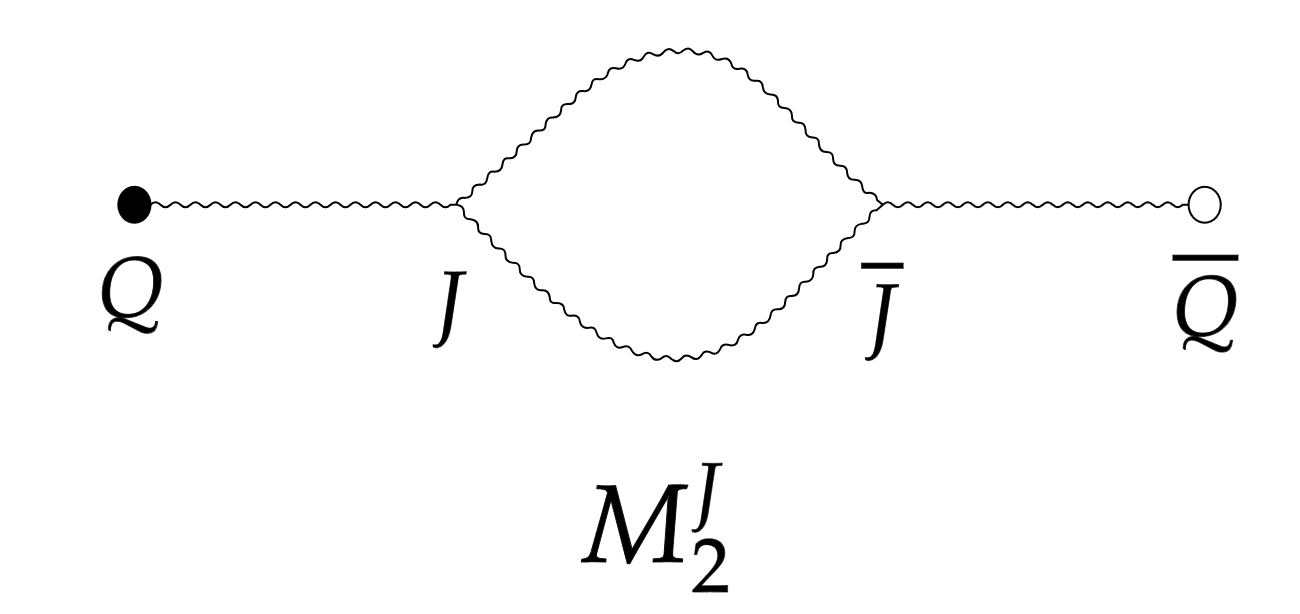} 
\includegraphics[scale=0.11]{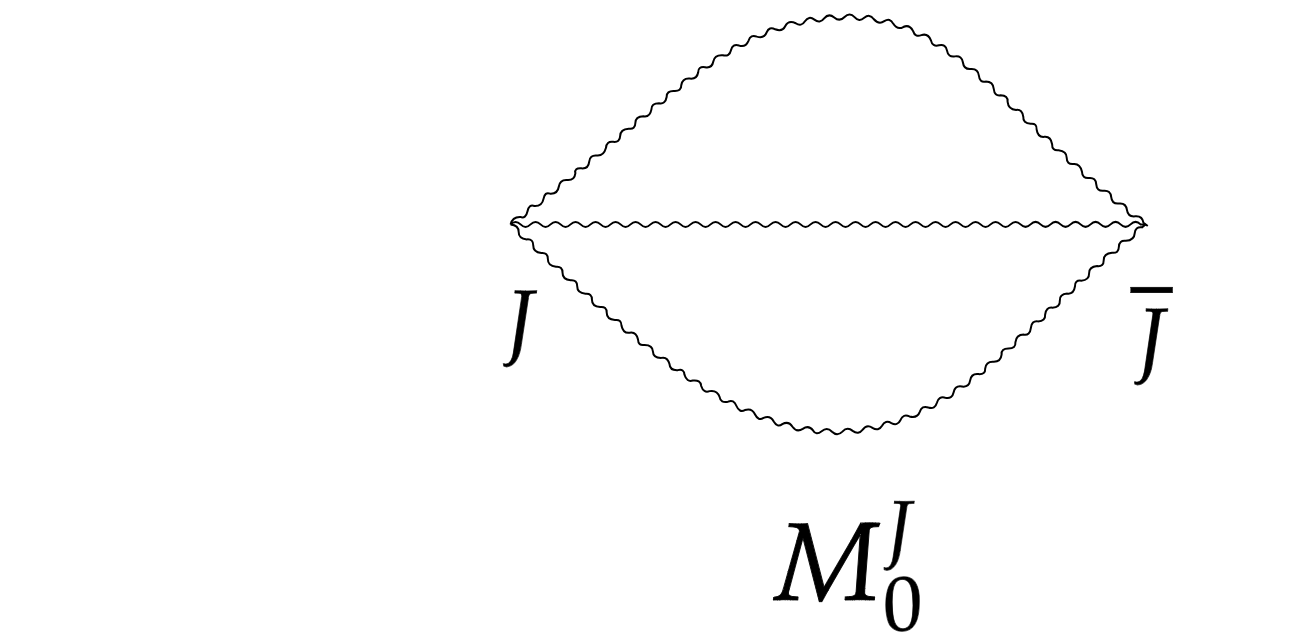} 
\caption{(Top) Semi-inclusive processes in which the baryon junction mediates the forward baryon production. Either 2,1 or 0 valence quarks can accompany the junction. The final-state mesons $X$ are produced as a result of 1, 2 or 3 strings fragmentation, respectively. Solid lines represent quarks, dashed lines represent the junction. (Bottom) Exotic meson states exchanged in the $t$-channel  of the corresponding process shown in the top panel. Wiggly lines represent gluonic strings.}
\label{fig:inclusive_amp}
\end{figure*}

At first glance, it may seem that testing gluonic baryon junctions in electron interactions (that couple to quarks rather than gluons) is not a good idea. We will show that this is not the case. On the opposite, at different energies and rapidities, one may access not only the purely gluonic junction $t$-channel exchange, but also exchanges of the junction accompanied by one and two quarks, see the top panel of Fig. \ref{fig:inclusive_amp}. In Regge theory description, these processes are described by Regge trajectories of $M_0^J=(J{\bar J})$, $M_2^J= (Q{\bar Q}J{\bar J})$, and $M_4^J= (QQ{\bar Q}{\bar Q}J{\bar J})$ exotic mesons \cite{Rossi:1977cy}, see the bottom panel of Fig. \ref{fig:inclusive_amp}. 
The corresponding intercepts, as discussed below, are $\alpha^J_0 \simeq 1/2$, $\alpha^J_2 \simeq 0$, and $\alpha^J_4 \simeq -1/2$.

Because the exchanges with largest intercepts dominate at high energies and rapidity differences, we predict that the production of baryons in the forward rapidity region (current fragmentation) at high energy will be dominated by the purely gluonic junction exchange (the rightmost diagram in Fig. \ref{fig:inclusive_amp}). As a result, the forward baryon will be uncorrelated in flavor with the target proton, and (modulo strange quark mass effects) the $p, \Lambda, \Xi$ and $\Omega$ baryons should be produced with equal probabilities. 

The dominance of gluonic junction exchange in this process predicted by Regge theory is easy to understand intuitively. Indeed, consider the forward baryon production process in $ep$ scattering in the center-of-mass frame. The valence quarks inside the target proton carry a large fraction of the proton's momentum, and are difficult to turn around (it would take a high momentum transfer photon-quark scattering with a small cross section). However the quark-antiquark component of the virtual photon's wave function can interact with the baryon junction inside the target proton and pull it in the forward region, see the rightmost diagram in the top panel of Fig. \ref{fig:inclusive_amp}. 
\vskip0.3cm

\begin{figure}
\includegraphics[scale=0.15]{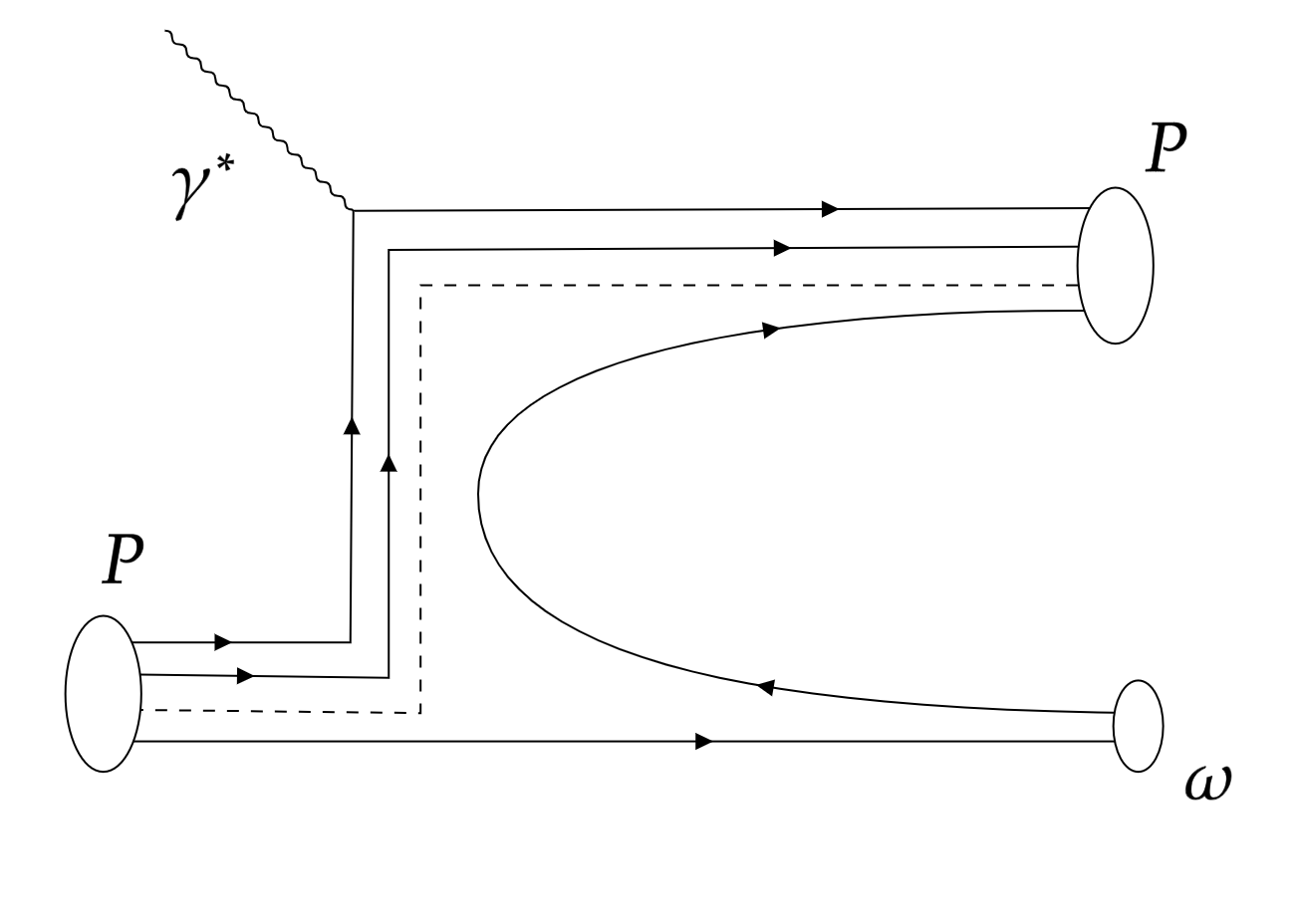} 
\caption{Exclusive process in which a proton is transferred to the photon fragmentation region while a $\omega$-meson is produced in the proton fragmentation region. Solid lines represent quarks, dashed line represents the junction.}
\label{fig:exclusive}
\end{figure}

The motivation for our study comes from a recent  exclusive $\omega$-production experiment at JLab where it was found that in a substantial fraction of events the proton is transferred to the forward region (with $\omega$ meson produced in the backward region)  \cite{Li:2017xcf}, see Fig. \ref{fig:exclusive}. However, because the process is exclusive, one has to exchange the entire baryon in the $t$-channel, and this amplitude is not sensitive to the presence of the junction. In other words, one cannot separate the flow of valence quarks from the flow of baryon number, which is the signature of baryon junction in high energy interactions. Such separation requires semi-inclusive electron-proton interactions which are the subject of this study.

\vskip0.3cm

\begin{figure}
\includegraphics[scale=0.2]{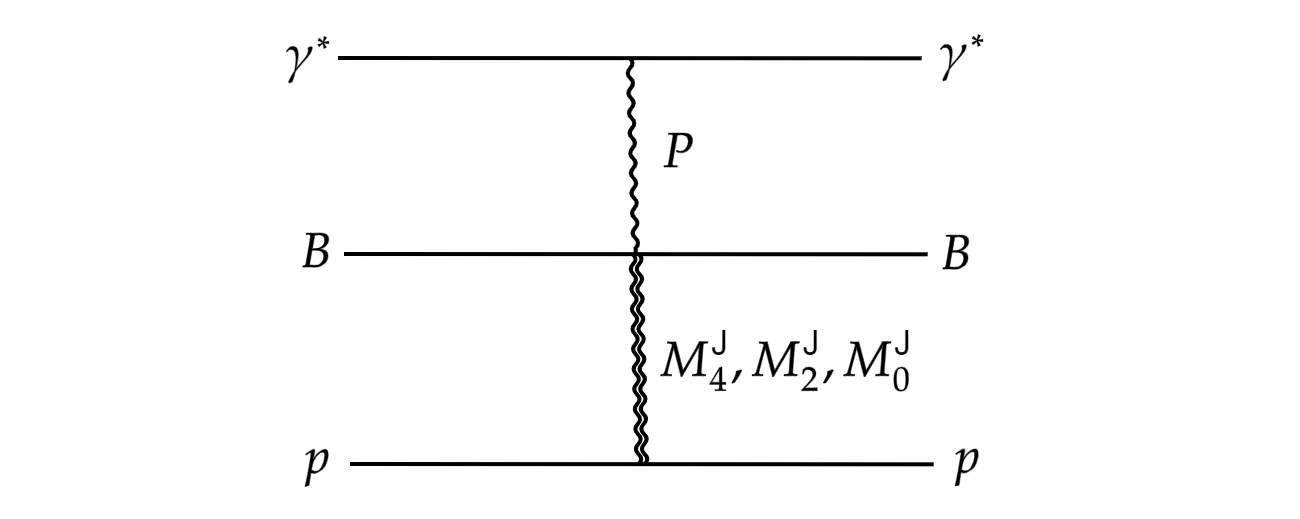}
\caption{3-particle elastic scattering process with a pomeron and exotic mesons exchanged in the $t$-channels of sub-diagrams.}
\label{fig:forward_amp}
\end{figure}

To estimate the energy dependence of the total cross section, we use the generalization of the optical theorem to semi-inclusive processes  developed by Mueller and Kancheli \cite{Mueller:1970fa, Kancheli:1970gt}. According to the Mueller-Kancheli theorem, the semi-inclusive cross section for the process of interest $\gamma^* + p \rightarrow B + X$ is related to the imaginary part of the forward scattering amplitude $\gamma^* + p + \bar{B} \rightarrow \gamma^* + p + \bar{B}$. In Regge theory such a process can be mediated by an exchange of Pomeron between $\gamma^*$ and $B$ and an exchange of baryonium between $p$ and $B$, as shown in Fig.\ref{fig:forward_amp}. The baryonium exchanged can be a $M_4^J$ state  (with $QQ\bar{Q}\bar{Q}J\bar{J}$ content), a 
$M_2^J$ ($Q\bar{Q}J\bar{J}$) state or a $M_0^J$ ($J\bar{J}$) state; these exotic mesons are shown in the bottom panel of Fig. \ref{fig:inclusive_amp}.

\vskip0.3cm

The energy and rapidity dependence of each of these processes can be computed in Regge theory. It is well known that the amplitude of a high-energy process depends on the square of center-of-mass energy $s$ (assumed to be much larger than all other energy scale in the process) as 
\beq
    \mathcal{A}\propto s^{\alpha(t)},
\eeq
where $\alpha(t)$ is the Regge trajectory of the states exchanged in the $t$-channel. 
In the forward scattering amplitude we set $t=0$. Introduce the momenta of $\gamma^*$ and $p$ in the center-of-mass (c.m.s.) frame: $p_1 = (\dfrac{\sqrt{s}}{2} , \dfrac{\sqrt{s}}{2}, 0, 0)$ and $p_2 = (\dfrac{\sqrt{s}}{2} , -\dfrac{\sqrt{s}}{2}, 0, 0)$, respectively; the momentum of the produced baryon is $p_B = (m_t \cosh y^* , m_t \sinh y^*, p_B^\perp)$. Here $s = (p_1+p_2)^2$, $m_t$ is the transverse mass of the produced baryon and $y^*$ is its c.m.s. frame  rapidity. Then
\begin{align}
s_1 = (p_1 + p_B)^2 = \sqrt{s}\, m_t\, e^{-y^*},  \\
s_2 = (p_2 + p_B)^2 = \sqrt{s}\, m_t\, e^{y^*},
\end{align}
and from the Mueller-Kancheli theorem
\beq
    \frac{d^3\sigma}{dp_B^3} \propto s_1^{\alpha_P(0)-1} s_2^{\alpha_M(0)-1} .
\eeq

The intercepts of  baryonium states were studied in \cite{Rossi:1977cy}. By using topological expansion and Regge factorization one can obtain
\begin{align}
    \alpha_4^J(0) = 2\alpha_B - 1 + (1-\alpha_R(0))\approx -\frac{1}{2}, \\
    \alpha_2^J(0) = 2\alpha_B - 1 + 2(1-\alpha_R(0))\approx 0, \\
    \alpha_0^J(0) = 2\alpha_B - 1 + 3(1-\alpha_R(0))\approx \frac{1}{2},
\end{align}
where the value of the baryon intercept has been taken equal to $\alpha_B = 0$ and the meson Regge intercept $\alpha_R = \frac{1}{2}$. Since the purely gluonic $M_0^J$ state has the largest intercept, the corresponding exchange should dominate in the high-energy limit:
\beq \label{eq:prediction}
\frac{d^3\sigma}{dp_B^3} \propto s^{-1/4} e^{-y^*/2},
\eeq 
where for the sake of an estimate we set the (soft) pomeron intercept equal to 1.
\vskip0.3cm

What are the observational consequences of the dominant $M_0^J$ exchange? First of all, in such a process all of the valence quarks in a baryon are replaced by other quarks from virtual quark-antiquark pairs. This means that the quark content of the baryon emerging in the forward direction could be arbitrary, with equal probabilities for different quark flavors (modulo the strange quark mass effect). Based on this, we expect the same number of $p, n, \Lambda, \Xi, \Omega$ baryons produced in semi-inclusive measurements tagging a baryon in the photon fragmentation region.

This should be contrasted to the case where $M_2^J$ or $M_4^J$ are exchanged. In the first case one of the valence quarks from the target proton remains in the final-state baryon so $\Omega$ cannot be produced. In the case of $M_4^J$ exchange two valence quarks remain in the final-state baryon and therefore neither $\Xi$ nor $\Omega$ can be produced. However, at sufficiently high energy and/or rapidity, the $M_0^J$ exchange dominates, and the forward baryon production should become flavor-blind.

\begin{figure}
\includegraphics[scale=0.5]{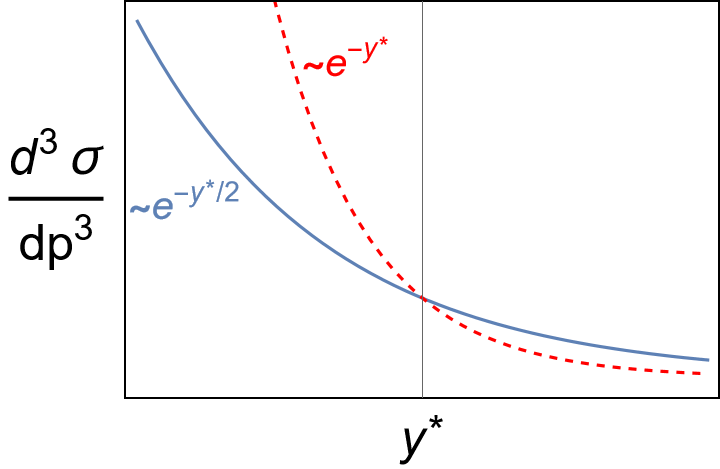}
\caption{The baryon junction prediction for the c.m.s. rapidity distribution of forward baryons is shown in blue. Dashed red line indicates a cross section  expected in conventional model.}
\label{fig:rapidity}
\end{figure}

Eq. (\ref{eq:prediction}) provides a prediction for the rapidity distribution of produced baryons at midrapidity (in the c.m.s. frame). This dependence is plotted in Fig. \ref{fig:rapidity} in comparison to the faster falling off cross section expected in a conventional model with $\alpha_B=0$, i.e. without the baryon junction. Therefore, we suggest to measure the rapidity distribution of the produced forward baryons. Proportionality to $e^{-y^*/2}$ would be a strong evidence in favor of the baryon junction transfer mechanism described above.

Finally, consider multiplicity of the produced mesons. Similarly to the case of hadron collisions, this multiplicity is correlated with the presence of the stopped baryon \cite{Kharzeev:1996sq}. Indeed, for the dominant process with $M_0^J$ exchange the final state mesons are produced from three fragmenting strings. In the first approximation (large $N$) the strings fragment independently, so we expect the multiplicity of mesons to be three times the multiplicity produced in a single string fragmentation, independent of the produced baryon. In the case of $M_2^J$ exchange, there are two fragmenting strings, and in the case of $M_4^J$ there is only one. 
Because a typical inelastic event at high energy is described by the Pomeron exchange, and Pomeron in topological expansion corresponds to the exchange of a closed string, the corresponding average multiplicity ${\bar n}$ results from the fragmentation of two strings. Therefore the multiplicities observed in various mechanisms of forward baryon production are related to the multiplicity ${\bar n}$ of an inclusive event by
\begin{align}
    n(M_0^J) = 3 n(M_4^J) &= \frac{3}{2}\ {\bar n}, \\
    n(M_2^J) &= {\bar n}, 
    \\
    n(M_4^J) &= \frac{1}{2}\ {\bar n}.
\end{align}
We therefore propose to measure both the rapidity dependence of forward baryon production and the associated multiplicity. 
\vskip0.3cm
To summarize, the energy, rapidity, and flavor dependence of forward baryon production in electron-proton interactions at JLab and EIC can further establish whether gluons can carry baryon number in QCD.
\vskip0.3cm

We thank Nicole Lewis, Prithwish Tribedy and Zhangbu Xu for useful discussions, and Abhay Deshpande for support and interest in this study. This work was supported by the U.S. Department of Energy under Grants DE-FG88ER41450 (DF, DK), DE-SC0012704 (DK) and DE-FG02-05ER41372 (WL). 
\bibliographystyle{ieeetr}
\bibliography{main.bib}

\end{document}